\documentclass[conference]{IEEEtran}
\IEEEoverridecommandlockouts
\usepackage{cite}
\usepackage{amsmath,amssymb,amsfonts}
\usepackage{algorithmic}
\usepackage{graphicx}
\usepackage{textcomp}
\usepackage{xcolor}
\usepackage{footmisc}
\usepackage{multirow}
\usepackage{threeparttable}
\usepackage{leftidx}
\def\BibTeX{{\rm B\kern-.05em{\sc i\kern-.025em b}\kern-.08em
    T\kern-.1667em\lower.7ex\hbox{E}\kern-.125emX}}
\begin{document}

\title{{\fontsize{20}{20}\selectfont Do Sports and Politics Mix? Cross-Analysis of Fan Bases of Major League Sports and Presidential Candidates}}

\author{\IEEEauthorblockN{
         Shuaidong Pan\IEEEauthorrefmark{1},
         Faner Lin\IEEEauthorrefmark{1},
         and Jiebo Luo\IEEEauthorrefmark{2}
        }
     \IEEEauthorblockA{\IEEEauthorrefmark{1} Goergen Institute for Data Science, University of Rochester, Rochester, NY, USA}
     \IEEEauthorblockA{\IEEEauthorrefmark{2} Computer Science Department, University of Rochester, Rochester, NY, USA}
     \IEEEauthorblockA{\IEEEauthorrefmark{1}\{span11, flin11\}@u.rochester.edu}
     \IEEEauthorblockA{\IEEEauthorrefmark{2}jluo@cs.rochester.edu}}

\maketitle

\begin{abstract}

Considering the fact that sports and politics interact in a very complex way, this interdisciplinary area remains largely untouched in data science research. Given the fact that huge fan bases exist for the major sports leagues such as NBA and NFL, it would be important for us to understand the hidden relationship between sports fans and their political preferences, and how do these preferences affect their behaviors in supporting different candidates during the presidential election. Taking advantage of the rich user data from Twitter, we propose a new metric, Congressional Devotedness Score, to model candidate preferences more accurately. 
Using the proposed metric, a fine-grained analysis is conducted at sport-level, state-level, and team-level for fans with strong political affiliation. While some of the findings conform to previous studies and reports, we also offer newer insights and quantitative evidences for all the findings. 

\end{abstract}

\begin{IEEEkeywords}
Sports, Presidential Election, Social Media, Following Behavior.
\end{IEEEkeywords}

\section{Introduction}

Sports and politics are always considered as two different domains, and numerous studies have been done in both fields from many perspectives. However, sports teams and political parties share many similarities in nature, and both are considered as expressions of a supporter's sense of self. Both types of supporters form communities that share the same value in many aspects and have a strong preference in their stances. In recent years, sports and politics become increasingly intertwined with each other, generating a ripple effect on both political supporters and sports fan groups. 

Previous studies have investigated the sports and their fan bases from sociology and social psychology perspectives. These studies illustrate how fandom can influence identity creation and community forming\cite{jacobson1979social},\cite{zhang2010buyer}. However, the relations between the fan bases of sports and American politics still remain unclear. With the wide-spread use of online social networks for both sports events and political campaigns, the microblogging service becomes a major platform to gather supporters and share messages. In our study, we adopt Twitter as our main tool to analyze the followers of the two major sports leagues in the United States, namely NBA and NFL, and the current presidential candidates as of early April 2020, namely Donald Trump, Joe Biden, and Bernie Sanders. Of primary interest are two simple but largely unanswered questions: 1) How to measure a user's presidential candidate preference accurately when the social media posts or public opinions of a user are not known; 2) what are the behavioral patterns of NBA and NFL fans with respect to following the presidential candidates.

We notice that there exist limitations when only analyzing the following relationship between sports fans and presidential candidates, and using such a method does not exclude the swing voters and casual followers. To examine the candidate preference among users with a relatively strong political preference and political interests, we first introduce a new metric, Congressional Devotedness Score. This metric allows an accurate representation of one's candidate preference, especially when the posts and text data are not available. Based on the scores, we further study the candidate preferences of NBA and NFL fan bases aggregated by sports, state, and team, and the results will be presented in the following sections.

\section{Related Work}

Previous works have studied the growing effects of international politics in sports, such as Olympics and other sporting events,  stating that sports can be a very useful diplomatic tool in international politics \cite{riordan2002international},\cite{strenk1979price}. Other empirical analyses from sports journalists pointed out that the depth of the involvement between domestic sports and American politics also affects the public’s political preference \cite{SportsPolitics},\cite{NFLPolitics}. However, in these analyses, most of the data is collected through surveys with a relatively small sample size. Using such a method not only requires a considerable amount of work but also would inevitably suffer from biases and noise. As we mentioned earlier, given the prominent role that Twitter plays in both presidential election and sporting events, and given the large user base of Twitter, we think it would be a more suitable data source and  more representative for studying the behaviors of sports fans during the presidential election. 

Inspired by the work of \cite{wangyu2017follow}, \cite{ACM2011computing} where relationships between celebrities’ followers or media outlets’ followers and their twitter following behavior are investigated, this  work aims at analyzing the political preference of sports fan bases. Based on the result of the analysis, we introduce the concept of Congressional Devotedness Score to measure the users' preference of the presidential candidates with respect to their interests in politics. Moreover, this method can be employed in studying the collective behavior for other special interest groups.
\begin{table}[t]
\caption{Followers Numbers of Selected NBA and NFL Teams}
\begin{center}
\begin{tabular}{llrlr}
\hline
\hline\\[-0.8em]
  \textbf{States} & \textbf{NBA Teams} & \textbf{\#Followers} & \textbf{NFL Teams} & \textbf{\#Followers}\\
  \hline\\[-0.8em]
  NY & \begin{tabular}{@{}l@{}}Nets \\ Knicks\end{tabular} &\begin{tabular}{@{}l@{}}1.05 \\ 2.16\end{tabular} &
  \begin{tabular}{@{}l@{}}Bills \\ Jets$^{1}$\\Giants$^{1}$\end{tabular}&\begin{tabular}{@{}l@{}}1.05 \\ 1.25\\1.86\end{tabular}\\\\[-0.8em]
  \hline\\[-0.8em]
  CA & \begin{tabular}{@{}l@{}}Warriors \\Clippers\\Lakers\\Kings\end{tabular} &\begin{tabular}{@{}l@{}}6.40\\1.56\\8.28\\1.07\end{tabular} & 
    \begin{tabular}{@{}l@{}}Rams$^{2}$ \\ Chargers \\Raiders \\49ers \end{tabular}&\begin{tabular}{@{}l@{}}0.89\\0.87\\1.66\\2.17\end{tabular}\\\\[-0.8em]
 \hline\\[-0.8em]
  OH & Cavaliers & 3.25 & 
    \begin{tabular}{@{}l@{}}Browns \\ Bengals \end{tabular}&\begin{tabular}{@{}l@{}}1.30\\0.85\end{tabular}\\\\[-0.8em]
\hline\\[-0.8em]
  FL & \begin{tabular}{@{}l@{}} Heat \\ Magic\end{tabular} &\begin{tabular}{@{}l@{}}4.69\\1.50\end{tabular} & 
    \begin{tabular}{@{}l@{}}Jaguars \\ Dolphins \\ Buccaneers \end{tabular}&\begin{tabular}{@{}l@{}}0.68\\1.02\\0.79\end{tabular}\\\\[-0.8em]
\hline\\[-0.8em]
  TX & \begin{tabular}{@{}l@{}}Mavericks \\ Rockets \\ Spurs \end{tabular} &\begin{tabular}{@{}l@{}}1.67\\3.00\\4.07\end{tabular} & 
    \begin{tabular}{@{}l@{}}Cowboys \\ Texans \end{tabular}&\begin{tabular}{@{}l@{}}3.93\\1.93\end{tabular}\\\\[-0.8em]
\hline\\[-0.8em]
  GA & Hawks & 1.26 & 
    Falcons & 2.38\\\\[-0.8em]
    \hline
    \hline\\[-0.8em]
\end{tabular}
\begin{tablenotes}\footnotesize
      \item Note: Number of followers is in million\\
      \item 1: Although Jets and Giants are located in New Jersey, their fan bases are mainly from New York
      \item 2: Rams moved from Saint Louis to Los Angeles in 2015
\end{tablenotes}

\end{center}
\label{tab1}
\vspace{-0.2cm}
\end{table}
 
\section{Data Acquisition}
Our dataset was collected in April 2020 with three major components using the Twitter Developer API\footnote{developer.twitter.com}. The first part of our data contains the Twitter follower ID information from the three presidential candidates in April 2020: Donald Trump, Bernie Sanders, and Joe Biden. We notice that Bernie Sanders has dropped out of the race after our study has been conducted, but it does not affect the users' behaviors in the time period that our paper primarily focuses on.

The second component of our data contains the follower IDs from the selected NBA and NFL teams. In order to make our analysis more equitable, we choose NBA and NFL teams from six states, New York, California, Ohio, Florida, Texas, and Georgia. New York and California are known to be reliable Democratic states (``Blue States") whereas Texas and Georgia are the states carried by Republicans (``Red States") in the four most recent elections. Ohio and Florida are two swing states that are targeted by both Democrats and Republicans in the election. The respective NBA and NFL teams in those selected states and the number of followers are reported in Table \ref{tab1}. In total, we have 25 million unique users following the 13 selected NBA teams and 15 million users following the 14 selected NFL teams. 

The last component of our data comprises the exhaustive follower ID information from the 100 senators of the 116th Congress of the United States (we exclude Bernie Sanders who is a major presidential candidate in the race), of which there are 53 Republicans, 45 Democrats, and 2 Independents who both caucus with Democrats \footnote{https://www.congress.gov/members}. With the senator follower information, we will be able to take the political interest into consideration when analyzing the candidate preferences for the NBA and NFL teams.



\section{Analysis and Result}
In this section, we report on: (1) the candidate following patterns of the selected NBA and NFL teams, (2) the new metric of Congressional Devotedness Score, (3) the application of our metric to analyzing the  candidate following patterns. 

\begin{table}[t!]
\caption{State-Level Following Ratios for Three Candidates}
\vspace{-0.15cm}
\begin{center}
\begin{tabular}{lllllll}
\hline\hline
States & \multicolumn{2}{c}{Sanders} & \multicolumn{2}{c}{Biden} & \multicolumn{2}{c}{Trump} \\
       & NBA & \multicolumn{1}{r}{NFL} & NBA  & \multicolumn{1}{r}{NFL}  & NBA  & \multicolumn{1}{r}{NFL} \\
CA     &0.222&0.255&0.132&0.187&\textbf{0.647}&\textbf{0.557}\\
NY     &0.233&0.225&0.250&0.242&\textbf{0.518}&\textbf{0.533}\\
OH     &0.180&0.185&0.187&0.197&\textbf{0.632}&\textbf{0.618}\\
FL     &0.232&0.244&0.221&0.203&\textbf{0.546}&\textbf{0.553}\\
TX     &0.215&0.178&0.149&0.156&\textbf{0.636}&\textbf{0.666}\\
GA     &0.216&0.196&0.209&0.167&\textbf{0.574}&\textbf{0.637}\\

\hline\hline\\[-0.8 em]
\end{tabular}
\end{center}
\label{state_diff_1}
\vspace{-0.4cm}
\end{table}

\begin{table}[t]
\caption{Congressional Devotedness Ratios for Candidates at State-level}
\vspace{-0.15cm}
\begin{center}
\begin{tabular}{lllllll}
\hline\hline
States & \multicolumn{2}{c}{Sanders} & \multicolumn{2}{c}{Biden} & \multicolumn{2}{c}{Trump } \\
       & NBA & \multicolumn{1}{r}{NFL} & NBA  & \multicolumn{1}{r}{NFL}  & NBA  & \multicolumn{1}{r}{NFL} \\
CA     &0.376&0.353&\textbf{0.386}&\textbf{0.416}&0.238&0.231\\
NY     &0.340&0.329&\textbf{0.519}&\textbf{0.489}&0.141&0.181\\
OH     &0.288&0.269&\textbf{0.457}&\textbf{0.412}&0.254&0.318\\
FL     &0.288&0.279&\textbf{0.426}&\textbf{0.404}&0.285&0.317\\
TX     &0.288&0.259&0.339&0.319&\textbf{0.373}&\textbf{0.422}\\
GA     &0.308&0.294&\textbf{0.401}&0.349&0.291&\textbf{0.357}\\

\hline\hline\\[-0.8 em]
\label{state_diff}
\end{tabular}
\end{center}
\vspace{-0.4cm}
\end{table}

\subsection{NBA and NFL Team Following Pattern}
\begin{figure}[t]
      \begin{minipage}[h]{1.0\linewidth}
        \centerline{\includegraphics[width=9cm, height=6cm]{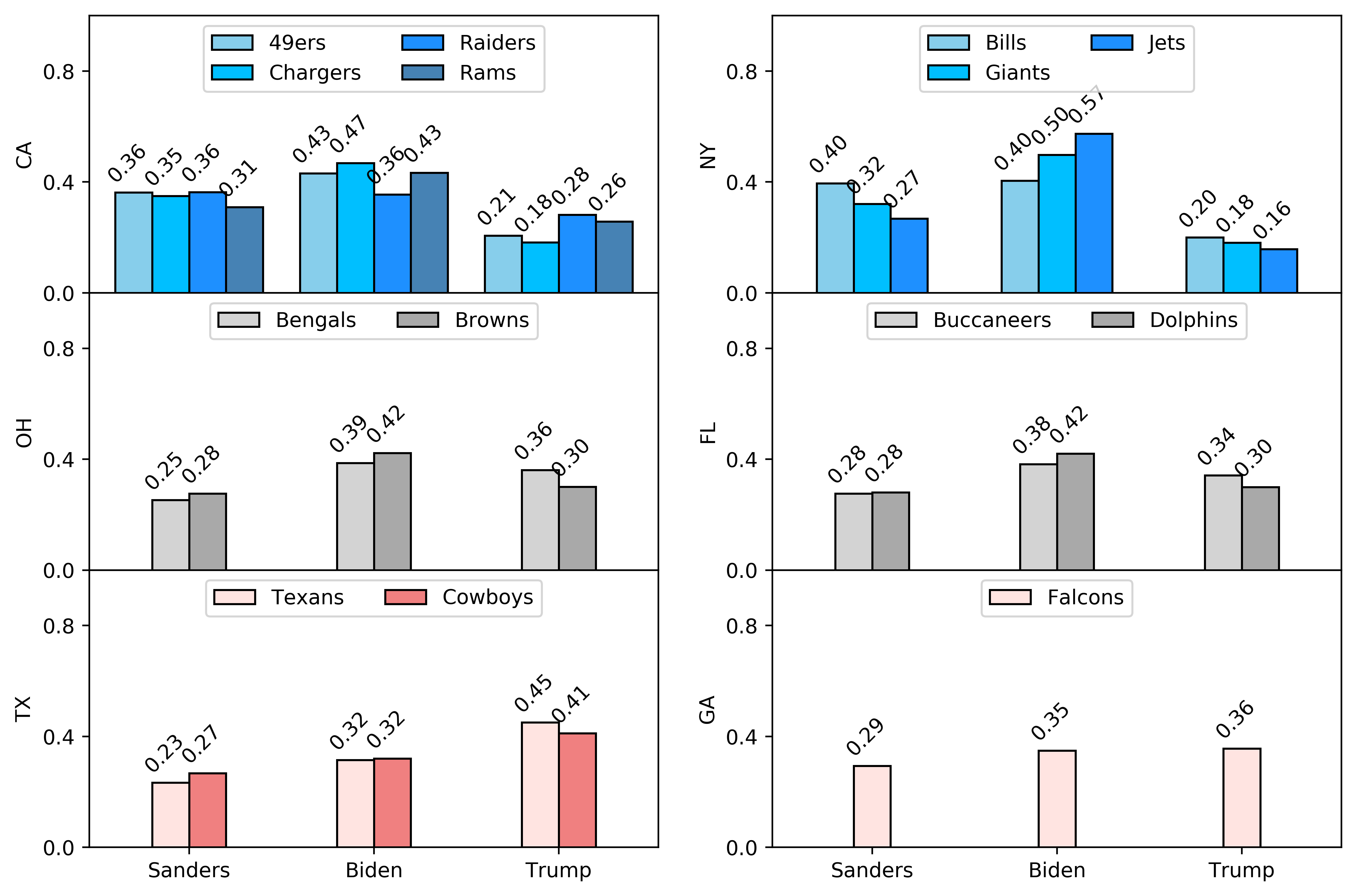}}
      \end{minipage}
\caption{Congressional Devotedness Ratios for NFL Teams}
\label{fig1}
\vspace{-0.1cm}
\end{figure}

According to the followers’ data collected in April 2020, it is found that Trump is the dominant force with the highest number of followers on  Twitter, 76.9  million, followed by Bernie Sanders of 11 million followers and Joe Biden of 4.7 million followers. 
In our study, we examine the candidate following patterns within a more specific context, i.e., selected teams from the NBA and the NFL. For the purpose of this study, we remove users that follow several NBA or NFL teams in order to filter out unusual  followers, such as journalists. After filtering, we have around 19 million unique users for 13 selected NBA teams and 12 million unique users for 15 selected NFL teams. Next, we examine the candidate following rates of both NBA and NFL teams, and find that the NBA has an average rate of 9.3\% whereas the NFL has a rate of 10.9\%. The ratio difference implies that the NFL fans on Twitter are more interested in the presidential election compared to the NBA fans on Twitter. In total, approximately 1.7 million (8.9\%) NBA fans and 1.3 million (10.7\%) NFL fans  follow at least one of the three candidates.

We observe from Table \ref{state_diff_1} that among NBA and NFL followers across the six states who are also following presidential candidates, the majority of the users are following Trump. A high number of total followers of Trump could be a result of Trump's status as the incumbent president and a proportion of Trump followers are fake and random followers \cite{bilton2016trump}.

To provide a more precise measure of the teams' candidate preferences, we include the followers' ID information of 100 senators of 116th Congress, and it is reasonable to believe that following senators from a specific Party is an indicator of a user's interest in politics. By collecting the followers from the senators' official Twitter accounts, we obtain 11,317,498 unique users on Twitter following Democrat senators and 9,453,552 unique users following Republican senators. We reselect 1,245,036 (6.5\%) NBA fans and 1,220,785 (9.6\%) NFL fans following at least one senator from our original dataset, and these users are considered to have a strong interest in politics than users who do not follow any senator. 

\begin{table}[htp]
\begin{center}
\caption{Senator Following Ratio of NBA and NFL}
\begin{tabular}{lrr}
\hline\hline
                            & NBA   & NFL   \\
                            \hline
Following Democrat Senators & 0.518 & 0.472 \\
Following Republican Senators & 0.309 & 0.358 \\
Following Senators From Both Parties & 0.173 & 0.170 \\
\hline\hline
\label{tableIV}
\end{tabular}
\end{center}
\end{table}

Next, we examine the distributions of both Democrat and Republican senators followers in both NFL and NBA fan bases. It is found that around 80\% of the senator followers only follow senators from one Party. We can see from Table \ref{tableIV} that for both NBA and NFL, the ratios for following Democrat senators only are the largest, compared to following Republican senators only and following senators from both parties. 
However, there exist some differences between the NBA and NFL fans. Fans of NBA are more discriminating and have a higher ratio of following Democrat senators, and this difference is smaller in NFL but also significant (12\%). The finding is consistent with the literature and anecdotal evidence that the NBA is more popular in the Democrat markets whereas the NFL has the least partisan fan base among popular sports leagues \cite{NFLPolitics}.

\subsection{Congressional Devotedness Score}

After analyzing the senator following distributions, it is reasonable for us to believe that a large proportion of Trump's followers are not strongly interested in politics. We notice that although Trump has a high following rate among NBA and NFL fans on Twitter, it is surprising that the Republican senators' following ratio is much lower than expected among the users with strong political interest. Since we aim to analyze the candidate preference for the followers with strong political interests,  simply looking at candidate following ratios would not be an accurate measurement. To solve this problem, a new metric, Congressional Devotedness Score (CDS) is proposed, and this score can help us measure the candidate preferences of each user with respect to their preference for a certain party.

For each follower of the three candidates, we compute the Congressional Weight for the two parties:
\begin{equation}
  Weight_{Party}\hspace{0.1cm}_i =
    \begin{cases}
      \frac{\alpha_i} {(\alpha_i+ \beta_i)} & \textit{if Party $=$ Democrats}\\
      \frac{\beta_i} {(\alpha_i+ \beta_i) } & \textit{if Party $=$ Republicans}
    \end{cases}       
\end{equation}
where $\alpha_i$ is the number of Democrat senators $user_i$ followed and $\beta_i$ is the number of Republican senators $user_i$ followed. Meanwhile, the Devotedness Score measures how committed a user is to a certain candidate. For a user following both Trump and Biden, the devotedness score will be 1/2, and for a user followering only one candidate, the devotedness score will be 1. Therefore, we combine the Congressional Weight and Devotedness Score to calculate the Congressional Devotedness score of $user_i$ and $candidate_j$:

\begin{equation}
{CDS_j}=\sum_{i} (Weight_{Party}\hspace{0.1cm}_i*\frac{\sigma_{ij}} {\sum_{j} (\sigma_{ij})})
\end{equation}
where $\sigma_{ij}$ equals 1 if $user_i$ follows $candidate_j$, and $\sigma_{ij}$ equals 0 otherwise. 

\begin{figure}[t]

      \begin{minipage}[h]{1.0\linewidth}
        \centerline{\includegraphics[width=9cm, height=6cm]{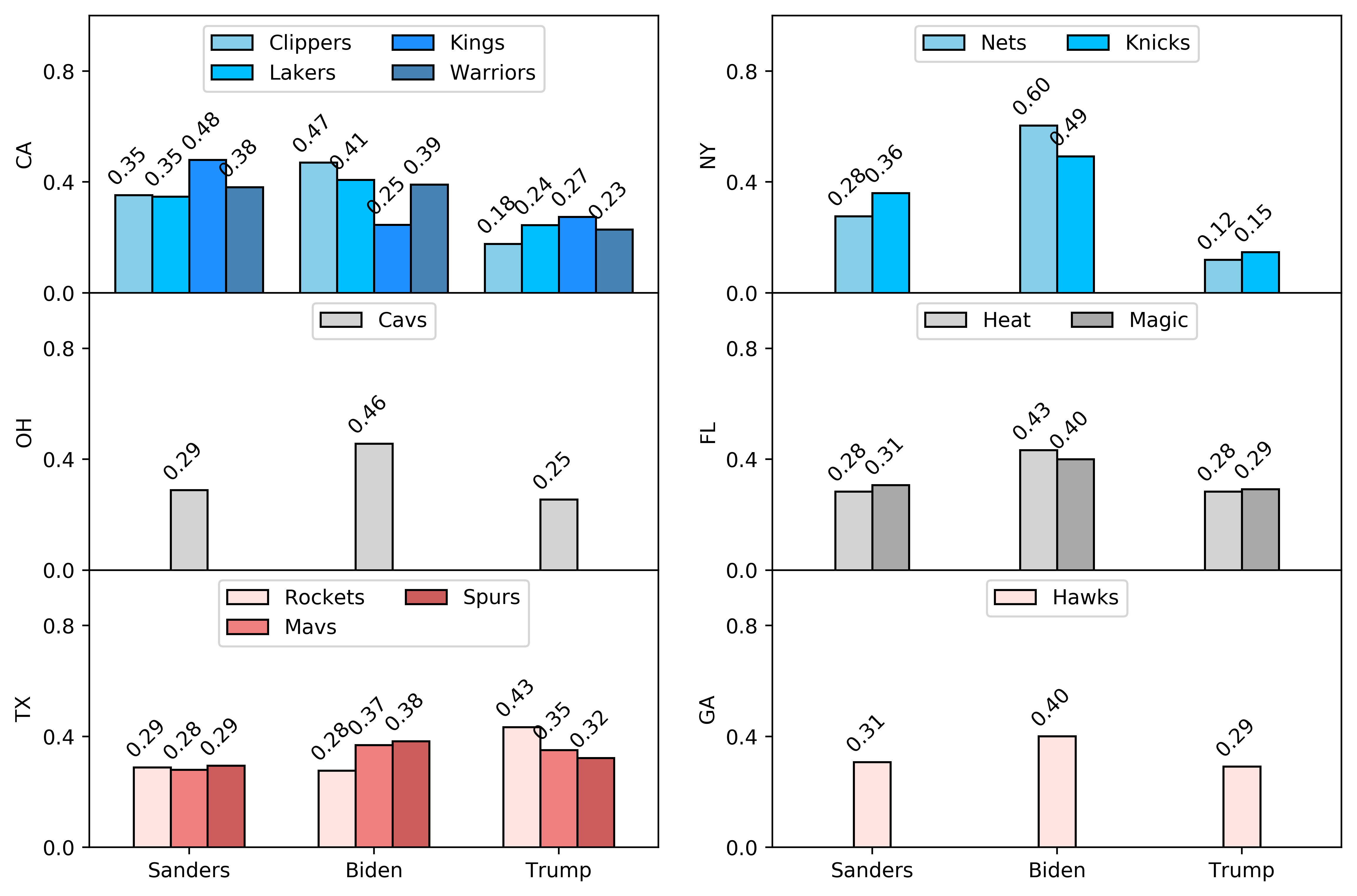}}
      \end{minipage}
\caption{Congressional Devotedness Ratios for NBA Teams} 
\vspace{-0.1cm}
\label{fig2}
\end{figure}

\subsection{Comparisons of Sport-level, State-level, and Team-level Congressional Devotedness Ratios}

In the previous section, we introduced the Congressional Devoteness Scores (CDS) for measuring the candidate preferences of users regardin their political interests and preferences. In this section, we further calculate the CDS for all users who follow at least one senator and one candidate. We then use this score to analyze the candidate preferences at  three levels: sport-level, state-level, and team-level. 

To compare the preferences of the  candidates across sports, states and teams, we introduce Congressional Devotedness Ratio (CDR), which is simply the weight of a candidate with respect to the sum of CDS of all candidates. 

\begin{table}[htp]
\vspace{-0.25cm}
\caption{Congressional Devotedness Ratios for Candidates at Sport-level}
\vspace{-0.25cm}
\begin{center}
\begin{tabular}{lrrr}
\hline\hline
Sports & Sanders  & Biden  & Trump  \\
\hline
NBA      & 0.335       & 0.404       & 0.261\\

NFL      & 0.302       & 0.398       & 0.300\\
\hline\hline
\label{NFL_NBA_compaire}
\end{tabular}
\end{center}
\vspace{-0.5cm}
\end{table}

 Table \ref{NFL_NBA_compaire} records the sport-level CDRs for the three candidates. Unlike in Section A, where Trump is the dominant force among all three candidates, we observe that now Trump's CDRs are the lowest while the CDRs for Biden are the highest. NBA also has a lower Trump CDR compared to NFL and the lower ratio for Republican candidates is consistent with our previous findings that NBA followers follow more Democrat senators than Republican senators. 
 
We also study the alignment and misalignment between the political preference of states and the candidate preference of teams through the state-level CDRs. In Table \ref{state_diff}, teams from the Democratic states, California and New York, have the lowest CDRs of Trump among all the six states, and Texas, the republican state, has the highest CDRs of Trump for both Sports. Moreover, the other Republican state, Georgia, has a higher CDR for Trump in NFL. Such a  distribution aligns with the states' political stances: 
although the sum of CDRs for the two Democrat candidates is always higher than the CDR for Trump overall, teams from the Blue States lean more towards the Democrat candidates and teams from the Red States lean more towards the  Republican candidate. 
Meanwhile, for all the states across the two sports, Biden's CDRs are generally higher than those of Sanders (the scores for Biden are the highest in New York). The high CDRs of Biden among Democrat followers are consistent with the Super Tuesday outcome, where Biden won 10 out of 14 states 
\cite{NYtimesSuper}. 

The Congressional Devotedness Ratios at team-level are presented in Figure \ref{fig1} and Figure \ref{fig2}. It is note worthy that {\it even the teams of the same sports in the same state can have different CDRs for the three candidates.} 
For the three NFL teams in New York, the CDRs for Sanders and Biden differ as Sanders has a higher supporting rate from the Buffalo Bills but a relatively low supporting rate from the New York Jets. This is likely due to the geographic difference in the fan bases of the two teams: the fans for the Buffalo Bills are  mainly located in the Western New York whereas the fans for the New York Jets are mainly located around  the New York City\footnote{https://interactive.twitter.com/nfl\_followers2014}. We observe a similar discrepancy in the ratios for Trump in the NBA teams of Texas: Trump's ratio is the highest for the Houston Rockets (43\%) while the lowest for the San Antonio Spurs (32\%). After checking the fan bases of Spurs and Rockets, we find that the followers of Rockets are mainly scattered around Houston and Eastern Texas whereas the followers of Spurs are mainly located around San Antonio and in Southern Texas. Overlaying the fan bases with the past election results\footnote{https://www.sos.state.tx.us/elections/historical}, we find that a considerable proportion of Spurs' fans is located in the precincts that voted for Democrats in the 2016 presidential election, whereas the fans of Rockets are concentrated in the areas where Republican is more popular.

\section{Conclusion and Future Work}
We obtain the Twitter followers ID information from three different groups: three presidential candidates still running in early April 2020, teams from two major sports leagues, and the 100 senators from 116th Congress. A new metric, Congressional Devotedness Score, is proposed to focus on users with strong political interest and measure their degrees of candidate preferences. Based on this metric, candidate preferences of the NBA and NFL are analyzed and investigated at the sport-level, state-level, and team-level. Our findings suggest: 1) at the sport-level, a large proportion of NBA and NFL fans with strong political interests lean towards Democrats; 2) at the state-level, the aggregated teams' political preference is aligned with the state's political stance; and 3) at the team-level, teams from same state exhibit  different candidate preferences mainly due to the geographic or demographic  differences that lead to the regional voting preferences. While some of our findings conform to the broad understanding, we also offer newer insights 
at the sport-level and team-level, as well as quantitative evidences for all the findings. 

\bibliographystyle{IEEEtran}
\bibliography{ref}

\end{document}